\documentstyle[11pt,newpasp,twoside,epsf]{article}
\def\edcomment#1{\iffalse\marginpar{\raggedright\sl#1\/}\else\relax\fi}
\reversemarginpar

\begin{document}
\title{Cosmological Microlensing
}
 \author{Joachim Wambsganss}
\affil{Universit\"at Potsdam, 
	Institut f\"ur Physik, 
	Am Neuen Palais 10, 
	14067 Potsdam, 
	Germany\\ 
and \\ 
	Max-Planck-Institut f\"ur Gravitationsphysik, 
	``Albert-Einstein-Institut",
	Am M\"uhlenberg 1,
	14476 Golm,
	Germany}

\begin{abstract}

Variability in gravitationally lensed quasars  can be due  to 
intrinsic fluctuations of the quasar or due to ``microlensing"
by compact objects along the line of sight. 
If disentangled from each other, microlens-induced
variability can be used to study two cosmological issues of
great interest,  
the size and brightness profile of quasars on one hand,
and the distribution of compact (dark)  matter along the line of sight.
In particular, multi-waveband observations are useful for this goal.

In this review recent theoretical progress as well
as observational evidence for quasar microlensing  
over the last few years will be summarized. 
Comparison with numerical simulations  will show ``where we stand".
Particular emphasis will be given to the questions microlensing
can address  regarding the
search for  dark
matter, both in the halos of lensing galaxies and in a
cosmologically distributed form.
A discussion
of desired observations and required theoretical studies
will be given  as a conclusion/outlook.

\end{abstract}

\section{What is cosmological microlensing?}

\subsection{Mass, length and time scales}
The lensing effects of  cosmologically distant 
compact objects in the mass range 
$$  10^{-6} \le m/M_{\odot} \le 10^6$$
on background objects is usually called ``cosmological microlensing".
The ``source" is typically a background quasar, but in principle other
distant source can be microlensed as well, i.e. distant supernovae or
gamma-ray bursters.  The only ``condition" is that the source size
is comparable to or smaller than the Einstein radius of the respective
lenses.

The microlenses can be ordinary stars, brown dwarfs, planets, black holes,
molecular clouds, or other compact mass concentrations (as long as their
physical size is smaller than the Einstein radius).
In most practical cases, the micro-lenses are part of a galaxy which 
acts as the main (macro-) lens.  However, microlenses could also
be located in, say, clusters of galaxies or they could even be 
imagined ``free floating" and filling intergalactic space.

The relevant length scale for microlensing is the Einstein radius of the lens: 
	$$ r_E = 
\sqrt{ { {4 G M } \over {c^2} } { {D_S D_{LS} \over D_L}  } }
\approx 
	4 \times 10^{16} \sqrt{M / M_\odot} \rm \,  cm, $$
where ``typical" lens and source redshifts of $z_L \approx 0.5$  and
$z_S \approx 2.0$  were assumed for the expression
on the right hand side ($G$ and $c$ are the gravitational
constant and the velocity of light, respectively; $M$ is the mass of
the lens, $D_L$, $D_S$, and $D_{LS}$ are the angular diameter distances
between observer -- lens, observer -- source, and lens -- source, respectively).

This length scale translates into an angular scale of
	$$ \theta_E = r_E/D_S  
		\approx  10^{-6} \sqrt {M /M_\odot} \ \ \rm arcsec. $$
It is obvious that the image splittings on these angular
scales  can not be observed directly. What makes microlensing observable
anyway is the fact that observer, lens(es) and source move relative to each
other. Due to this relative motion, the micro-image configuration 
changes with time, and so does the total magnification, i.e. the
sum  of the magnifications of all the micro-images. And this change
in magnification over time can be measured:
microlensing is a ``dynamical" phenomenon. 

There are two time scales involved: the standard lensing
time scale $t_E$ is the time it takes the source to cross the
Einstein radius of the lens, i.e. 

	$$ t_E = r_E/v_\perp   \approx  15 \sqrt {M / M_\odot}  v_{600}^{-1} \ \ \rm  years, $$
where the same typical assumptions are made as above, 
and the relative transverse velocity $v_{600}$ is parametrized in 
units of 600 km/sec. 
This time scale results in            discouragingly large values for
stellar mass objects. 
However, we can expect fluctations on much shorter time intervals. Due
to the fact that the magnification distribution is highly non-linear, the
sharp caustic lines separate regions of low and high magnification. So
if a source crosses such a caustic line, we will observe a large
change in magnification during the time it takes the source to
cross its own diameter: 

	$$ t_{cross} 
= R_{source}/v_\perp   \approx   4 R_{15}  v_{600}^{-1} \ \ \rm  months.$$
Here the quasar size $R_{15}$ is parametrized in units of
$10^{15}$cm. This time scale $t_{cross}$ can be significantly shorter
than $t_E$.

\subsection{Geometry }
The typical geometry of a cosmological microlensing situation is
displayed in Figure 1.  
The main lensing agent is a
galaxy, which consists partly of stars (and other compact objects).
The ``macro"-lensing situation produces two (or more) images of the background
quasar, separated by an angle of order one arcsecond. However, because of the
graininess of the main lens, each of these macro-images consists of many
micro-images, which are separated by angles of order microarcseconds, and
hence are unresolvable. However, due to the relative motion between source,
lens and observer, the micro-image configuration changes, and so
does the total magnification, which is observable.

\begin{figure}
\plotone{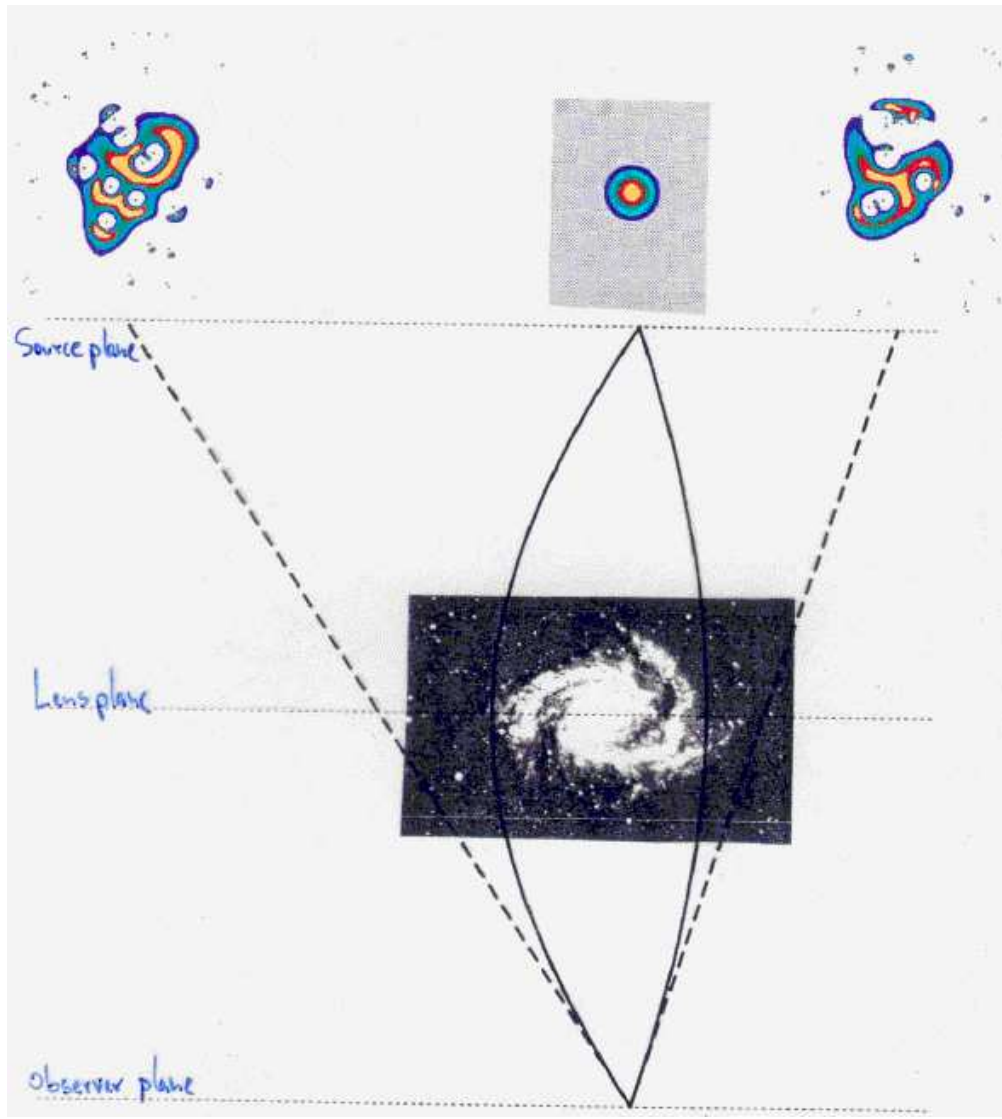}
\caption{\protect{\label{fig-geo}} Geometry of a microlensing situation: the galaxy
in the lens plane acts as a ``macro-lens", producing two separate images
of the background quasar. Due to the graininess of the matter distribution
in the galaxy, each macro-image is split into many micro-images. Only the
total magnification of all the microimages is measurable. }
\end{figure}

\subsection{History }
Only a few months after the detection of the first multiply imaged
quasar was published by Walsh, Carswell and Weymann (1979), 
Kyongae Chang and Sjur Refsdal suggested in their paper ``Flux variations of QSO 0957+561 A, B and image splitting by stars near the light path",
(Chang \& Refsdal 1979):

\begin{quote}
\small
``If the double quasar QSO 0957+561 A, B is the result of gravitational
lens actions by a massive galaxy, stars in its outer parts and close 
to the line of paths may cause significant flux changes in one year."
\end{quote}

\noindent
Only 
two years later, J. Richard Gott III asked the question: ``Are heavy halos
made of low mass stars? A gravitational lens test", 
Gott (1981). He suggests that a heavy halo made of low mass stars in the range
$4 \times 10^{-4} M_\odot \ \ \ {\rm to } \ \ \ 0.1 M_\odot$
\begin{quote}
\small
`` ... 
should produce fluctuations of order unity in the intensities of the QSO 
images on time scales of 1-14 years."
\end{quote}
He went on to propose:
\begin{quote}
\small
``Observations of QSO 0957+561 A, B and other quasars over time can establish
whether the majority of mass in the heavy halo is in the form of low mass
stars." 
\end{quote}

\noindent
In a number of further papers, the lensing effect of
individual stars on the background quasar was explored; e.g.,
Young (1981) did some numerical simulations and 
applied them to the double quasar. 
In the year 1989, the first observational evidence for
quasar microlensing was presented: 
Irwin et al. (1989) showed that fluctuations in image A of the
quadruple quasar Q2237+0305 could not be due
to intrinsic variability of the quasar. 
Such fluctuations could be explained by the lensing action of
low-mass main sequence stars and allowed conclusions on the
quasar size to be of order a few times $10^{14}$ cm 
(Wambsganss, Paczy\'nski,  \& Schneider 1990).

\subsection{Early Promises}

Fluctuations in the brightness of a quasar can have two causes: they
can be intrinsic to the quasar, or they can be microlens-induced.
For a  single quasar image, the difference is hard to tell. However,
once there are two or more gravitationally lensed (macro-)images of a 
quasar, we have a relatively good handle to distinguish the two
possible causes of variability: any fluctuations due to
intrinsic variability of the quasar have to show up in all
the quasar images, after a certain time delay. In fact, time delays
of quasars are only measurable {\it because} quasars are variable intrinsically.
(This argument could even be turned around: the measured time delays
in multiple quasars are the ultimate proof of the intrinsic
variability of quasars.)  So once a time delay is measured in a multiply-imaged
quasar system, the incoherent fluctuations can be contributed to microlensing.

There is another possibility to distinguish the two causes of fluctuations:
even without measuring the time delay, it is possible to tell whether
measured fluctuations are intrinsic or not: in some quadruple lens systems,
the image arrangement is so symmetrical around the lens, that any
possible lens model predicts very short time delays (of order days or
shorter), so that fluctuations in individual images that are 
longer than the (theoretical) time delay and not followed by corresponding
fluctuations in the other images, can be safely attributed to microlensing.
This is in fact the case of the quadruple system Q2237+0305,
see below.

Early on, the papers exploring microlensing 
made four predictions concerning the possible scientific successes. 
With microlensing
we should  be able to 
\begin{enumerate}
\item
	determine the effects of compact objects between 
		the observer and the source,
\item
	determine the size of quasars,
\item
	determine the two-dimensional brightness profile of quasars,
\item
	determine the masses of lensing objects.
\end{enumerate}
In Chapter 3 the observational results to date will be discussed in 
some detail.  It can be stated already here that 1) has been achieved. 
Some limits on the size of quasars have been obtained, so 2) is partly
fulfilled. We are still (far) away from solving promise 3), and 
concerning point 4) it is fair to say that it was shown  
that the observational results are consistent with 
masses of the lensing objects corresponding to low-mass stars.

\begin{table}[thb]
\begin{center}
\begin{tabular}{|c||c|c|}
\hline\hline
&&\\
Lensing galaxy:  & Milky Way & Lens in Q0957+561 \\
&&\\
\hline\hline
&&\\
distance to Macho? & no & yes \\
&&\\
velocity of Macho? & no & (no) \\
&&\\
 mass?   &  ???   &  ???   \\
&&\\
 optical depth?  & $ \approx 10^{-6}$  & $\approx$ 1  \\
&&\\
Einstein angle (1 M$_\odot$)? & $\approx$ 1 milliarcsec  & $\approx$ 1 microarcsec  \\
&&\\
time scale?  & hours to years  & weeks to decades \\
&&\\
event? & individual/simple & coherent/complicated \\
&&\\
default light curve?      & smooth & sharp caustic crossing \\
&&\\
when/who proposed? & Paczy\'nski 1986 & Gott 1981 \\
&&\\
first detection?   & EROS/MACHO/OGLE & Irwin et al. 1989 \\
		   &                 1993 & \\
&&\\
\hline\hline
\end{tabular}
\caption{The important lensing properties for the
two regimes of microlensing -- local group vs. cosmological -- 
are compared to each other. At the left the various properties
of interest are named, in the middle and right-hand column it is listed
whether these properties are known and/or the 
corresponding values are given for the Milky Way and the lensing galaxy,
respectively.}
\end{center}
\end{table}

\subsection{``Local Group" Microlensing versus Extragalactic Microlensing}

This contribution deals mainly with quasar 
microlensing, where in most cases the surface mass density (or optical depth)
is of order the critical one. 
In contrast to that, 
most other papers at these proceedings are concerned with 
the ``local group" or low optical  depth
microlensing. Since there are a number of similarities, but as well quite some
differences between these two regimes, in Table 1 the
various quantities are compared to each other.

\section{Theoretical Work on Cosmological Microlensing}
In the situation of a multiply imaged quasar, the surface mass density (or
``optical depth") at the position of an image is of order unity. If this
matter is made of compact objects in the range described above, due to the
relative motion of source, lens(es) and observer,  
microlensing is expected to be going on basically ``all the time". In addition,
this means that the lens action is due to a coherent effect of many microlenses,
because the action of two or more point lenses whose projected positions
is of order their Einstein radii adds in a very non-linear way. 

The lens action of more than two point lenses cannot be easily treated
analytically any more. Hence numerical techniques were developed in order
to simulate the gravitational lens effect of many compact objects.
Paczy\'nski (1986) had used a method to look for the extrema in the
time delay surface. Kayser, Refsdal, Stabell (1986), Schneider 
\& Weiss (1987) and Wambsganss (1990) had developed and applied
an inverse ray-shooting technique that
produced a two-dimensional magnification distribution in the source plane.
An alternative technique was proposed by Witt (1993) and Lewis et al. (1993).
They solved the lens equation along a linear source track.
All the recent theoretical work on microlensing is based on either
of these techniques.

Fluke \& Webster (1999) explored analytically  caustic crossing events for
a quasar. Lewis \& Belle (1998) showed that 
spectroscopic monitoring of multiple quasars can be used to probe the
broad line regions.
Wyithe et al. (2000a, 2000b) explored and found 
limits on the quasar size and on the 
mass function  in Q2237+0305.

Agol \& Krolik (1999)
and 
Mineshige \& Yonehara (1999) developed techniques to recover
the one-dimensional brightness profile of a quasar, based on the
earlier work by Grieger et al. (1988, 1991). Agol \& Krolik showed
that frequent monitoring  of a caustic crossing event in many
wave bands (they used of order 40 data points in eleven filters over
the whole electromagnetic range), one can recover 
a map of the  frequency-dependent brightness distribution
of a quasar!
Yonehara (1999) in a similar approach explored the effect of 
microlensing on two different accretion disk models.
In another paper, Yonehara  et al. (1998)
showed that monitoring  a microlensing event in 
X-rays can reveal structure of the quasar accretion
disk as small as AU-size.

With numerical simulations and limits obtained from three
years of Apache Point monitoring data of Q0957+561, and based
on the Schmidt \& Wambsganss (1998) analysis, we
extend the limits on the masses of ``Machos" in the (halo of the) lensing
galaxy: the small ``difference" between the time-shifted 
and magnitude-corrected lightcurves of images A and B 
excludes a halo of the lensing galaxy made of compact objects with
masses of $\le 10^{-2} M_\odot$ (Wambsganss et al. 2000).

\section{Observational Evidence for  Cosmological Microlensing}

\subsection{The Einstein Cross: Quadruple Quasar Q2237+0305}
In 1989 the first evidence for cosmological microlensing was
found by Irwin et al. (1989) in the quadruple quasar Q2237+0305:
one of the components showed fluctuations, whereas the others
stayed constant. In the mean time,
Q2237+0305 has been monitored by many groups
(Corrigan et al. 1991;  {\O}stensen et al. 1996;
Lewis et al. 1998; Wozniak et al. 1999).

The most recent and  most exciting results (Wozniak et al. 1999)
show that
all four images vary dramatically (but incoherently!), 
going up and down like
a rollercoaster in the last three years: 
\begin{itemize}
	\item $ \Delta m_A \approx$ 0.6 mag, 
	\item $ \Delta m_B \approx$ 0.4 mag, 
	\item $ \Delta m_C \approx$ 1.3 mag (and rising?), 
	\item $ \Delta m_D \approx$ 0.6 mag.
\end{itemize}
This is very encouraging news, and  it calls for continuing and 
expanding  monitoring programs for lensed quasars.

\subsection{The Double Quasar Q0957+561 }
The microlensing results for the double quasar Q0957+561 are 
at the moment not quite as exciting as those for Q2237+0305. 
In the first few years, there appears to
be an almost  linear change in the (time-shifted) brightness
ratio between the two images 
($ \Delta m_{AB} \approx 0.25$ mag over 5 years).
But since about 1991, this ratio
stayed more or less constant within about 0.05 mag, so 
not much microlensing was going on in this system recently
(Schild 1996; Pelt et al. 1998; Schmidt \& Wambsganss 1998;
Wambsganss et al. 2000).
At his moment, the possibility for some small amplitude rapid microlensing 
cannot be excluded; however, one needs a very well determined time delay
and very accurate photometry, in order to establish that (Colley \& Schild 1999).

\subsection{Other multiple quasars }
A number of other multiple 
quasar systems are being monitored more or less
regularly. For some of them microlensing has been suggested (e.g.
H1413+117,  {\O}stensen et al. 1997;  or B0218+357, Jackson et al. 2000)
In particular the possiblity for ``radio"-microlensing appears
very interesting (B1600+434,  Koopmans~\&~de~Bruyn~2000 and Koopmans,
these proceedings), because this was not expected in the first place, 
due to the presumably larger source size of the radio 
emission region. This novel aspect of microlensing
is definitely worth pursuing in more detail.

\section{Unconventional Microlensing}

\subsection{Microlensing in individual quasars? }
There were a number of papers interpreting the 
variability of individual quasars  as microlensing
(Hawkins 1996, 1998; Hawkins \& Taylor 1997). Although this
is an exciting possibility and it could help us detect a 
population of cosmologically distributed lenses, it is not
entirely clear at this point whether the observed fluctuations
can be fully or partly attributed to microlensing. After all, 
quasars {\it are}
intrinsically variable (otherwise we could not measure
time delays), and the amount of microlensing in
single quasars must me smaller than the one in 
multiply imaged ones, due to the lower surface mass density. 
Only more observations can help solve this question.

\subsection{``Astrometric Microlensing" Centroid shifts  }
An interesting aspect of microlensing was explored by 
Lewis \& Ibata (1998). They looked 
at centroid shifts of quasar images due to microlensing. At each 
caustic crossing, a new very bright image pair emerges or disappears,
giving rise to sudden changes in the ``center of light" positions.
The amplitude could be of order 100 microarcseconds or larger,
which should be observable with the next generation of astrometric 
satellites, like SIM.

\section{Microlensing: Now and Forever? }
Monitoring observations of various multiple 
quasar systems in the last decade have clearly established qualitatively
that the phenomenon of  ``cosmological"
microlensing exists. Uncorrelated variations in multiple quasar
systems with amplitudes of more than a magnitude have been
observed, on time scales
of weeks to months to years. However, in order to get close to a
really quantitative understanding, much better monitoring programs need
to be performed.

On the theoretical side, there are two important questions:
what do the lightcurves tell us about the lensing objects, and
what can we learn from them about the size and structure of the
quasar. As response to the first question,  
the numerical simulations are able to give a qualitative understanding
of the measured lightcurves (detections and non-detections), in general
consistent with ``conservative" assumptions about the object masses
and velocities. But
due to the large number of parameters (quasar size, masses of lensing
objects, transverse velocity) and due to the large variety of
lightcurve shapes possible even for a fixed set of parameters, 
no ``unique" quantitative 
explanation or even predictions has been achieved so far. 
The prospects of getting
much better sampled lightcurves of multiple quasars, as shown by the OGLE 
collaboration, should be enough motivation to explore this regime
more quantitatively in the future.

The question of the quasar structure deserves much more attention. Here
gravitational lensing is in the unique situation to be able to explore
an astrophysical field that is unattainable by any other means. Hence
much more effort should be put into attacking this problem. 
This involves much more ambitious observing programs, with the goal
to monitor caustic crossing events in many filters over the whole
electromagnetic spectrum, and to further develop numerical techniques
to obtain useful values for the quasar size and (one-dimensional) profile
from unevenly sampled data in (not enough) different filters.

Summarized it can be said that cosmological microlensing -- though 
still a young discipline  --
has already achieved  part of its original goals in attacking the
questions of compact (dark) matter and quasar size and structure. 
But there is still
a lot of very interesting astrophysics out there, ``in reach".
The field is definitely worthwhile pursuing with more efforts
in the future
 -- both theoretically and observationally.

\acknowledgments
It is a pleasure to thank the organisers, Penny Sackett and John Menzies
and their colleagues, for their  excellent 
macro- and micro-planning and running of 
``Microlensing 2000: A New Era of Microlensing Astrophysics".

\end{document}